\documentclass[10pt,preprint]{aastex}
\shorttitle{M~2-9, Mz~3, and He~2-104}
\shortauthors{Smith \& Gehrz}
\begin{document}

\title{BIPOLAR SYMBIOTIC PLANETARY NEBULAE IN THE THERMAL-IR:
M~2-9, Mz~3, and He~2-104}

\author{Nathan Smith\altaffilmark{1,2,3}}
\affil{Center for Astrophysics and Space Astronomy, University of
Colorado, 389 UCB, Boulder, CO 80309}

\and

\author{Robert D.\ Gehrz\altaffilmark{3}}
\affil{Astronomy Department, University of Minnesota, 116 Church St.\
SE, Minneapolis, MN 55455}

\altaffiltext{1}{Hubble Fellow; nathans@casa.colorado.edu}

\altaffiltext{2}{Visiting astronomer at the European Southern
Observatory, La Silla, Chile.}  

\altaffiltext{3}{Visiting astronomer at the IRTF, operated by the
University of Hawaii under contract with NASA.}  

\begin{abstract}

We present thermal-infrared images of three extreme bipolar objects,
M~2-9, Mz~3, and He~2-104.  They are bipolar planetary nebulae with
bright central stars and are thought to be powered by symbiotic binary
systems.  The mid-infrared images spatially resolve the spectral
energy distributions of the central engines from the surrounding
nebulae.  A warm dust component of several hundred degrees can account
for the core emission, while a cooler component of $\sim$100 K
produces the more extended emission from the bipolar lobes.  In every
case, the dust mass for the unresolved core region is orders of
magnitude less than that in the extended lobes, raising doubts that
the hypothetical disks in the core could have been responsible for
pinching the waists of the nebulae.  We find total masses of roughly
0.5--1 M$_{\sun}$ in the nebulae of M~2-9 and Mz~3, requiring that
this material was donated by intermediate-mass progenitor stars.  The
mass of He~2-104's nebula is much lower, and any extended emission is
too faint to detect in our images.  Extended dust emission is detected
around both M~2-9 and Mz~3, in both cases resembling the distribution
of ionized gas.  Our images of Mz~3 have the highest signal-to-noise
in the extended polar lobes, and we show that the fairly uniform color
temperature derived from our images can explain the 110~K dust
component that dominates the far-infrared spectral energy
distribution.  In the case of Mz~3, most of the mass traced by dust is
concentrated at high latitudes, and we note possible evidence for
grain destruction in shocks indicated by an anticorrelation between
[Fe~{\sc ii}] and dust emission.  Except for these regions with
enhanced [Fe~{\sc ii}] emission, the dust continuum closely resembles
the distribution of ionized gas.

\end{abstract}


\section{INTRODUCTION}

The formation of bipolar structure in planetary nebulae (PNe) is one
of the enduring challenges to understanding the late stages of stellar
evolution.  No simple model has yet emerged that can account for how a
diffuse slowly-rotating asymptotic giant branch star develops strong
bipolarity (not to mention point symmetry, multipolar outflows, jets,
etc.) as it becomes a PN, although confining circumstellar disks,
disk/jet precession, core rotation, or magnetic fields may play
important roles (reviews of shaping mechanisms are given by Balick \&
Frank 2002; Frank 1999; Mellema \& Frank 1995; and contributions in
Meixner et al.\ 2004).  One can alleviate the mental strain required
to invent self-consistent single-star models by invoking binary
systems to provide an axis of symmetry and angular momentum, which are
necessary for bipolar phenomena.

The ``Butterfly'' (M~2-9), the ``Ant'' (Mz~3), and the ``Southern
Crab'' (He~2-104) are among the most dramatic bipolar PNe with
tightly-pinched waists seen in {\it Hubble Space Telescope} ({\it
HST}) images (Balick \& Frank 2002; Balick 1999; Balick et al.\ 1997;
Corradi et al.\ 2001).  It is interesting that each of these is
thought to be powered by a symbiotic binary system (Whitelock 1987;
Balick 1989; Corradi \& Schwarz 1993; Corradi et al.\ 2000; Schmeja \&
Kimeswenger 2001; Smith 2003).  These objects have been studied
extensively with spectra at visual and near-infrared (IR) wavelengths.
They have bright but highly-reddened stellar cores with rich
emission-line spectra, accompanied by faint nebular spectra from lower
density regions in their polar lobes (Evans 1959; Cohen et al.\ 1978;
Allen \& Swings 1972; Swings \& Andrillat 1979; Balick 1989; Lutz et
al.\ 1989; Goodrich 1991; Hora \& Latter 1994; de Freitas Pacheco \&
Costa 1996; Phillips \& Cuesta 1999; Smith 2003); M~2-9 and Mz~3 in
particular are spectroscopic twins at visual and near-IR wavelengths
except for the lack of H$_2$ in Mz~3 (Smith 2003).  Each object
requires an ionization source with a temperature of order 30,000 K,
and the nebulae seem to have moderately-enhanced abundances of He and
N, suggesting intermediate-mass progenitors.  Published distances to
each object vary widely, scattered around 1-2 kpc; in this study we
quote numerical values like M and L for a nominal distance of 1 kpc.

By comparison, little is known about the nature of dust in these
sources.  The spatial distribution of thermal-IR emission can
constrain the dust temperature and mass in the core regions and polar
lobes separately, in order to then estimate important quantities like
the total mass of the nebulae and the mass of putative circumstellar
disks.  Low-resolution 8-13 $\micron$ spectra of M~2-9 and Mz~3 show
similar, nearly flat continuum emission in both sources, except for
the presence of [Ne~{\sc ii}] 12.8 $\micron$ in Mz~3 (Aitken \& Roche
1982), but these observations did not include any spatial
information. Although each object was observed with the {\it Infrared
Astronomical Satellite} ({\it IRAS}), the low spatial resolution of
those data cannot separate the relative contributions of the core and
polar lobes.  Ground-based images in the thermal-IR with
arcsecond-scale spatial resolution have only been published for Mz~3
(Quinn et al.\ 1996), but that was one image in a single broad filter,
making it difficult to constrain the dust temperature or mass.  In an
effort to clarify the spatial distribution of mid-IR properties of
M~2-9, Mz~3, and He~2-104, we obtained 8-25 $\micron$ images of these
sources with MIRLIN on the IRTF and TIMMI2 on the ESO 3.6m telescope.

\section{OBSERVATIONS}

\subsection{MIRLIN Images of M~2-9}

We obtained thermal-IR images of M~2-9 on the nights of 2001 July 22,
23, and 24 using the 5--26 $\micron$ camera MIRLIN mounted on the 3m
NASA Infrared Telescope Facility (IRTF).  In this configuration,
MIRLIN's 128$\times$128 Si:As BIB array has a pixel scale of
0$\farcs$465 and a field of view of 1$\arcmin$.  Chop-nod images were
obtained in three broadband filters, N1, Qs, and Q5 centered at 8.8,
17.9, and 24.5 $\micron$, respectively.  We also used the 2\%
resolution circular variable filter (CVF) centered at 10.5 and 12.8
$\micron$ to image the emission lines of [S~{\sc iv}] and [Ne~{\sc
ii}], respectively (see Table 1).  The weather conditions were not
optimal (intermittent thin clouds) on the first night when the 12.8
$\micron$ image was obtained, and the photometric uncertainty of that
image is $\pm$30\%, but the weather was better on the next two nights
when the other images were obtained, and for those the photometric
uncertainty is 10-20\% (larger uncertainties in the Q-band images).
The chopper throw was 30$\arcsec$ east/west with a similar north/south
telescope nod that we varied slightly from one set of images to the
next.  After sky subtraction, the many individual frames were
resampled to a smaller pixel scale and then shifted and added, using
the bright central star for spatial registration.  The observations
were flux calibrated using observations of $\alpha$ Sco, adopting the
zero magnitude fluxes in the MIRLIN handbook.

Figure 1 shows the final flux-calibrated images of M~2-9 in the N1,
Qs, and Q5 filters, where the lowest contour is drawn at approximately
3$\sigma$ above the background.  The [S~{\sc iv}] 10.5 $\micron$ and
[Ne~{\sc ii}] 12.8 $\micron$ images are not displayed in Figure 1
because only the central point source and no extended structure was
detected.  Even in the continuum images, the extended structure in the
nebula is very faint and only marginally detected, so we have not made
ratio maps displaying the color temperature and emitting optical
depth.  The faint mid-IR extended emission from warm dust is clearly
elongated in the north-south direction, similar to the structure seen
in {\it HST} and near-IR images (Balick \& Frank 2002; Hora \& Latter
1994).  This extended structure is discussed in more detail below in
\S 4.  Table 1 gives sky-subtracted flux densities for the central
star in each filter, measured in a 4$\arcsec$ diameter circular
software aperture.

\subsection{TIMMI2 Images of Mz~3 and He~2-104}

On 2003 May 15 we used the Thermal Infrared Multi-Mode Instrument
(TIMMI2) on the ESO 3.6m telescope at La Silla, Chile to obtain mid-IR
images of Mz~3 and He~2-104.  TIMMI2 has a 240$\times$340 pixel Si:As
BIB detector with a pixel scale of 0$\farcs$2 and a field of view of
48\arcsec$\times$64$\arcsec$.  Chop-nod images were obtained in three
broadband filters centered at 8.9, 11.9, and 17.0 $\micron$, and with
the [Ne~{\sc ii}] 12.8 $\micron$ narrowband filter (see Table 2).
Different chop-nod patterns were used for each source, due to the
different sizes of the nebulae in optical images.  The polar lobes of
Mz~3 have a spatial extent of roughly 20\arcsec$\times$30$\arcsec$ in
emission-line images (Lopez \& Meaburn 1983; Redman et al.\ 2000;
Smith 2003) allowing us to place the positive and negative images of
the nebula side-by-side on the array.  Therefore, to observe Mz~3 we
used a chopper throw of 40$\arcsec$ north-south (chopping off the
array) and east-west nods of 30--35$\arcsec$.  Since the inner bipolar
lobes or ``rings'' of He~2-104 are only $\sim$10$\arcsec$ across in
optical images (Corradi et al.\ 2001), we used a chop-nod pattern
appropriate for a point source, where we chopped 18$\arcsec$
north-south and nodded the telescope 20$\arcsec$ east-west.  This
allowed the target to be placed on the array four times for each set
of chop-nod observations, increasing the effective on-source exposure
time.  For flux calibration, we obtained similar observations of
$\gamma$~Cru on the same night and at similar airmass, using fluxes
for each filter listed in the TIMMI2 user manual.  The weather was
mostly photometric, and the photometric accuracy in our images is
5--10\%, dominated by uncertainty in the standard star calibration.

Figures 2 and 3 show the resulting TIMMI2 images of Mz~3 and He~2-104,
respectively, where the lowest contour is 3$\sigma$ above the
background.  Table 2 gives sky-subtracted flux densities for the
central star in each filter, measured in a 4$\arcsec$ diameter
circular software aperture for both Mz~3 and He~2-104.

Figure 2 reveals obvious extended thermal-IR emission from warm dust
in the bipolar lobes around Mz~3 in all three continuum filters, as
well as strong [Ne~{\sc ii}] emission.  The observed morphology is
consistent with the pair of polar bubbles along the nearly north-south
axis seen in emission-line images at shorter wavelengths (Smith 2003;
Santander-Garcia et al.\ 2004; Guerrero et al.\ 2004).  This extended
structure is discussed in \S 5.  Our images of He~2-104 had even
better sensitivity than those of Mz~3, but were not sensitive enough
to detect significant extended structure from the bipolar lobes or
rings of He~2-104.  The 11.9 and 12.8 $\micron$ images in Figure 3 do
show a very faint hint of extended structure, with the lowest contour
slightly elongated along a northeast/southwest direction out to
$\sim$3$\arcsec$ from the star.  Elongation along this direction
(which is different from both the chop and nod directions) would be
consistent with the elongated central emission structure seen in {\it
HST} images (Corradi et al.\ 2001), marking the limb-brightened
section near the equator where the two polar lobes appear to meet.
Our images demonstrate that the extended mid-IR structure of He~2-104
is much fainter compared to the central source than in either M~2-9 or
Mz~3.

\section{PHOTOMETRY OF THE CENTRAL ENGINES}

All three of our targets show a bright, unresolved central source at
mid-IR wavelengths, related to their status as potential symbiotic
binaries.  This is very different than the case of $\eta$ Carinae, to
which these three planetary nebulae are often compared, where
high-resolution IR images reveal a complex group of dust clumps
arranged in a disrupted torus at the point where the two polar lobes
meet at the equator (Smith et al.\ 2002).  Our images of both M~2-9
and Mz~3 reveal extended emission from warm dust in the bipolar lobes
of each object, while we detect no extended mid-IR emission around
He~2-104 at comparable sensitivity.  However, spatially-resolved
mid-IR photometry of these central engines reveals important clues to
the nature of all three objects.  The extended nebulae of M~2-9 and
Mz~3 are discussed separately in \S 4 and \S 5.

\subsection{Dust Temperature, Luminosity, and Mass}

Figure 4 displays the spectral energy distributions (SEDs) of the
bright central cores measured from our images using a 4\arcsec\
diameter aperture (from Tables 1 and 2), compared to the integrated
emission in a larger aperture over a wider range of IR wavelengths
measured by 2MASS, {\it MSX}, and {\it IRAS}.\footnote{These data were
collected from the Infrared Science Archive at {\url
http://irsa.ipac.caltech.edu/}.}  For each source, a solid curve shows
an approximate gray-body fit to the SED of the unresolved central
engine, which excludes extended emission from dust in the nebulae.  In
all three cases, the integrated emission at longer IR wavelengths
clearly exceeds this solid curve, and the dotted curve shows the
contribution from additional cool dust that must be emitted from a
larger area on the sky than the aperture we used to measure each of
the central stars (the dashed curve shows the sum of both, which fits
the total 8--100 $\micron$ SED for each object).  We take these two
dust temperature components to represent warm dust in the unresolved
central core and cool dust in the more extended nebulae.
Temperatures, luminosities, and dust masses for these warm and cool
dust components are listed in Table 3, calculated for a nominal
distance of 1 kpc.

The dust mass for each temperature component in Table 3 was derived
from the total integrated luminosity of the individual curves drawn in
Figure 4, with necessary assumptions about the grain properties.  The
relation for the mass of dust required to account for the observed IR
luminosity can be expressed as

\begin{equation}
M_d = \Big{[} \frac{a \rho}{3 \sigma Q_e T_d^4} \Big{]} L_d
\end{equation}

\noindent where $a$, $\rho$, and $Q_e$ are the effective grain radius,
mass density, and mean thermal emissivity of the optically important
grains, and $\sigma$ is the Stefan-Boltzmann constant.  Since no
silicate emission features are seen in any of the three objects we
observed, we assume that carbon grains dominate the IR emission, and
we adopt $\rho\simeq$2.25 g cm$^{-3}$.  We do not know the grain size,
but since the recurrent outbursts that may form symbiotic planetary
nebulae are likely caused by nova-like outbursts, we assume $a\la$0.2
$\micron$ (i.e., the maximum grain size observed in novae; Gehrz
1999).  With grains smaller than 0.2 $\micron$, the grain emissivities
given by Gilman (1974) can be approximated with

\begin{equation}
Q_e = \frac{1}{100} a T_d^2
\end{equation}

\noindent which permits us to write an expression for the dust mass
that is independent of the grain emissivity and radius.  We have

\begin{equation}
M_d = \Big{[} \frac{100 \rho}{3 \sigma T_d^6} \Big{]} L_d
\end{equation}

\noindent which is used to calculate the dust mass for each component
listed in Table 3 and discussed below for each object.  Given the
inherent uncertainties in the flux calibration and uniqueness of the
fits to the IR data, the dust masses we calculate are probably
uncertain at the $\pm$30\% level.  Table 3 also lists likely total
masses (gas + dust) assuming a gas-to-dust mass ratio of 230, which is
appropriate if the grains fully deplete the carbon in
solar-composition ejecta (Grevesse \& Anders 1989).

We presume that at shorter near-IR wavelengths, the continuum SED of
each object is dominated by the unresolved central source (e.g., Hora
\& Latter 1994; Smith 2003).  At those wavelengths, the SED is an
uncertain combination of reddened photospheric emission from the
central stars, hot dust in the core, and scattered light from the
lobes, so we do not show a fit to the 2MASS data in Figure 4.  Even if
extended hot dust in the polar lobes made a significant contribution
at these wavelengths, the total mass would be negligible compared to
the mass of cooler dust (e.g., Smith et al.\ 1998; and Table 3).  In
any case, the very simple fits in Figure 4 are sufficient to
characterize the dust emission that dominates in the mid-IR.

\subsection{Mz 3}

Of the three targets we observed, Mz~3 has the smallest fraction of
the total emission contributed by its unresolved core. At a wavelength
of $\sim$12 $\micron$, for example, the central source contributes
less than half the total flux measured by {\it MSX} and {\it IRAS}.
The relative contribution of the core weakens toward longer
wavelengths, supplying $\sim$10\% at 20 $\micron$, and only about 1\%
at 60 $\micron$.

The integrated IR luminosity is only about 25\% of the total presumed
bolometric luminosity of L=10$^4$ L$_{\odot}$ (Smith 2003), indicating
that the dust is optically thin to the escaping UV and
visual-wavelength stellar radiation over most of the solid angle seen
by the central engine.  The mass of dust in the central core is
difficult to guage accurately from the present data, because the
emission may be optically thick.  However, the dust mass needed to
emit the optically thin 320 K gray-body shown in Figure 4 gives a
useful lower estimate of the mass required in the outer part of the
disk.  

This core mass for the warm dust component is negligible compared to
the much larger total mass of $\sim$0.6 $M_{\sun}$ (gas + dust) for
the cooler material farther from the star in the bipolar lobes.  This
relatively large mass for the circumstellar ejecta, which approaches 1
$M_{\sun}$, reinforces the conjecture from chemical abundances that
the bipolar lobes were ejected from an intermediate-mass progenitor
star (Smith 2003).

Our imaging photometry reveals a significant 12.8 $\micron$ [Ne~{\sc
ii}] emission line from the central source, in agreement with spectra
taken by Aitken \& Roche (1982).  We measure a continuum-subtracted
flux of $\sim$9 Jy or 3.6$\times$10$^{-11}$ ergs s$^{-1}$ cm$^{-2}$
(integrated over the filter bandpass), and a corresponding equivalent
width of $\sim$530 \AA.

The observed mid-IR SED of the unresolved central source is somewhat
flatter than a single gray-body, implying the existence of a range of
dust temperatures.  Thus, the single temperature of 320 K is only
representative for the purpose of calculating the minimum luminosity
and mass of the dust (Table 3).  Indeed, an additional component of
hot dust at $\sim$900 K is required to fit the near-IR continuum
spectrum (Smith 2003; Cohen et al.\ 1978).  This wide range of
temperatures from the unresolved central source probably indicates the
existence of a circumstellar disk with dust located at a range of
different radii.

The observed properties of the core SED allow us to place some
constraints on the hypothetical disk.  The 320 K component represents
the cooler, outer parts of the disk, while the 900 K dust (Smith 2003)
traces the inner region of the disk, near the dust sublimation radius.
For L=10$^4$ L$_{\odot}$, these correspond to 10 and 80 AU,
repsectively, or 10 to 80 mas for D=1 kpc.  In any case the disk is
much smaller than the spatial resolution of our images and near the
limit attainable with {\it HST}, but within the current reach of IR
interferometry.  The disk is evidently very thin; the disk's IR
luminosity is less than about 5\% of the total available bolometric
luminosity (Smith 2003), indicating a half opening angle of
$\la$12$\arcdeg$.

\subsection{M 2-9}

The SED of M~2-9 shown in Figure 4$a$ is similar to that of Mz~3,
except that the relative contribution from extended cool dust is
weaker.  The 8--20 $\micron$ SED of the central source is almost
identical to Mz~3, both having an N-band flux of $\sim$30 Jy and an
intrinsic IR luminosity just below 500 L$_{\odot}$; however, M~2-9
lacks significant 12.8 $\micron$ [Ne~{\sc ii}] emission (Aitken \&
Roche 1982).  The 260 K fit to the SED of the central source is very
approximate, since the observed SED is actually flatter than any
single temperature, perhaps indicating dust at various radii in a
circumstellar disk.  As with Mz~3, a hotter dust component is needed
to account for the near-IR continuum (Hora \& Latter 1994).  The
central source also requires dust at a somewhat cooler average
temperature compared to Mz~3, implying that the central source is
intrinsically less luminous or that the dust is distributed toward
larger radii in the disk.

The cooler dust component that dominates at far-IR wavelengths,
emitted by the extended dust in the polar lobes of M~2-9, supplies a
smaller fraction of the total IR luminosity than its counterpart in
Mz~3.  Also like Mz~3, the warm dust in the core of M~2-9 makes a
negligible contribution to the total mass of the circumstellar nebula.
And finally, like Mz~3, the relatively large total mass we derive for
the cooler component in the extended nebula (about 0.8 $M_{\sun}$)
requires that the progenitor that ejected the mass was probably an
intermediate-mass star of at least a few solar masses.  

Altogether, the similarities between the IR properties of M~2-9 and
Mz~3 are remarkable.  Since H$_2$ formation is linked to dust grains
in most circumstances, these similarities underscore the mystery of
why the near-IR H$_2$ lines are so prominent in the polar lobes of
M~2-9 (Hora \& Latter 1994), while being totally absent in Mz~3 (Smith
2003).

\subsection{He 2-104}

All our targets are dominated by a bright unresolved core, but
He~2-104 is the only one in our sample showing no clear evidence for
extended emission at 8--12 $\micron$.  The bright unresolved central
engine of He~2-104 dominates the SED at all IR wavelengths we have
observed, and the core flux we measure matches the integrated 12
$\micron$ {\it IRAS} flux.  He~2-104 is also significantly fainter
than the other two sources; the warm dust component has less than 1/3
of the N-band flux or bolometric IR luminosity of M~2-9 or Mz~3.  The
SED of He~2-104 shows no evidence for any 12.8 $\micron$ [Ne~{\sc ii}]
emission.

In addition to the 320 K dust component that dominates at mid-IR
wavelengths, He~2-104 also requires a component of hotter dust to
account for the observed near-IR photometry, as is the case for both
M~2-9 and Mz~3.  Like the other two, the addition of a cooler dust
component is needed at far-IR wavelengths to explain the integrated
{\it IRAS} fluxes.  However, in He~2-104 the cool 95 K component is
far less luminous than the other two: it has only 30 L$_{\odot}$, and
it is the only source in which the luminosity of the cool dust
component at far-IR wavelengths is much lower than the warmer dust
component that dominates in the mid-IR.  From equation 3, the mass of
dust required to emit the observed far-IR flux from the 95 K component
in He~2-104 is only 10$^{-4}$ M$_{\odot}$, or a total gas + dust mass
of $\sim$0.02 $M_{\sun}$.  This is a much smaller mass than either
M~2-9 or Mz~3, which explains why we did not detect any extended
structure in our images, but does not let us place strong constraints
on the likely mass of the progenitor star.

\subsection{Where's the Donut?}

Here we briefly note that in all three objects, the mass of ejecta in
the unresolved central core/disk is more than 100 times lower than the
mass of ejecta in the more extended polar lobes.  This fact is
relevant for the quest to understand the origin of bipolar structure
(see Balick \& Frank 2002), since with disk masses that are orders of
magnitude less than the polar lobes, these disks cannot provide the
resistance needed to absorb momentum and thereby constrict the
equatorial expansion of an otherwise spherical shell.  The same
problem exists for $\eta$ Carinae as well, where the hot equatorial
dust in the core region provides negligible mass compared to the polar
lobes (Smith et al.\ 2003; Frank et al.\ 1998).  Some other mechanism
that favors polar ejection of material may be required (see e.g.,
Balick \& Frank 2002; Matt \& Balick 2004; contributions in Meixner et
al.\ 2004).  Observations of stars still on the asymptotic giant
branch suggest that asymmetries in proto-planetary nebulae start
early, either through rotation or binary interactions (Gehrz 2004).

\section{EXTENDED STRUCTURE OF M 2-9}

Figure 1 reveals clear extended structure in the bipolar lobes of
M~2-9, extending out to 15--20\arcsec\ from the star toward both the
north and south, consistent with the extent of the bipolar lobes seen
in {\it HST} and IR images (Balick \& Frank 2002; Hora \& Latter
1994).  Of the three wavelengths shown in Figure 1, the highest
quality image is Figure 1$b$ at 18 $\micron$. In this image, the right
(west) side of the nebula is clearly brighter than the left.  This is
the opposite of the case in emission-line tracers of ionized gas
(Balick \& Frank 2002; Hora \& Latter 1994), which are much brighter
on the east side of the nebula; note, however, that this pattern
changes with time (Doyle et al.\ 2000; Allen \& Swings 1972).  This
difference between emission lines and dust emission suggests a spatial
anticorrelation between ionized gas and warm dust in the variable
ionization structure of M~2-9's bipolar lobes.  Perhaps temporary
exposure to the hard UV radiation field of the central source is
sufficient to destroy significant quantities of dust.  The destruction
of dust grains by UV radiation is supported by observations of novae
(Gehrz et al.\ 1980a, 1980b).

In Figure 1, the polar lobes of M~2-9 appear somewhat thicker at 24.5
$\micron$ than they do at 8--18 $\micron$.  A larger size at longer IR
wavelengths might imply the existence of an outer shell of cooler dust
that emits less efficiently at shorter IR wavelengths.  Deeper images
with higher spatial resolution are obviously needed to confirm this
conjecture.  Although poorly justified by the quality of the images in
Figure 1, the possible existence of a double shell structure -- a cool
outer dust shell and a warmer inner dust shell -- is motivated by
other observational clues.  M~2-9 is frequently compared to the
Homunculus nebula around $\eta$ Car, which shows precicely this type
of double-shell structure in thermal-IR images (Smith et al.\ 2003).
A double-shell morphology is reinforced by the near-IR emission line
structure of $\eta$ Car, with H$_2$ delineating a thin outer shell,
and a smaller inner shell seen clearly in [Fe~{\sc ii}] lines (Smith
2002).  This is {\it identical} to the near-IR excitation structure
observed in these same emission lines in images of M~2-9 (Hora \&
Latter 1994).

In addition to the similarities in the dust temperature and nebular
excitation structure, M~2-9 and $\eta$ Car share other observational
characteristics.  Several observers have noted similarities in the
very rich emission-line spectra of the central sources, including
strong [Fe~{\sc ii}] emission (e.g., Balick 1989; Swings \& Andrillat
1979; Allen \& Swings 1972).  More recently, it has been recognized
that they also share a similar and very rare type of temporal
variability in their nebulae.  M~2-9 shows bizarre changes in its
apparent brightness distribution, which are usually attributed to an
azimuthally-evolving UV radiation field caused by moving shadows from
the cooler component in the central binary system (Doyle et al.\ 2000;
Allen \& Swings 1972).  Multiepoch UV images of $\eta$ Car have
recently revealed a similar type of variability in its ``Purple
Haze'', with an anlogous root cause (Smith et al.\ 2004).

Finally, Figure 1 (especially Fig.\ 1$b$) shows emission from the knot
S3, and possibly its northern counterpart N3 as well, both at roughly
15\arcsec\ from the star.  This is significant, because it may
indicate that dust survives in these condensations, which are thought
to be shock excited.  They are presumably formed by shocks in a fast
polar wind or jet, and are sometimes called FLIERS (fast
low-ionization emission regions; Balick \& Frank 2002) or ansae.
Their bright [Fe~{\sc ii}] emission is thought to be due to the
liberation of Fe into the gas phase following the destruction of dust.
Thus, their detection in dust emission at thermal-IR wavelengths is
significant.  However, [Fe~{\sc ii}] might contaminate the 17.9
$\micron$ filter.

\section{EXTENDED STRUCTURE OF Mz 3}

\subsection{Morphology in Images}

Of our three targets, Mz~3 shows the brightest and most dramatic
extended structure.  Although Mz~3 has a complex ejecta pattern with
multiple polar lobes seen in images (Smith 2003; Santander-Garcia et
al.\ 2004; Guerrero et al.\ 2004), here we are only concerned with the
innermost bipolar lobes (i.e. the head and abdomen of the ant).
Viewing the images on a computer display shows considerably more
detailed structure than is conveyed in Figure 2, so we applied 20
iterations of the {\sc lucy} deconvolution algorithm in IRAF to our
11.9 $\micron$ continuum and 12.8 $\micron$ [Ne~{\sc ii}] images,
which have the best signal-to-noise in our dataset.  The results are
shown in Figures 5$a$ and 5$b$, respectively.\footnote{The emission
spot immediately west of the central star in Figure 5$a$ is an
artifact of the sky subtraction offset settling time, as opposed to
real structure. Our standard star used as the PSF in the deconvolution
had a slightly different chop-nod pattern.}  At any position in the
polar lobes, the [Ne~{\sc ii}] intensity is significantly higher than
the continuum emission -- from Figure 5 we have typically
$I_{\lambda}$([Ne~{\sc
ii}])\,$\simeq$\,3\,$\times\,I_{\lambda}$(11.9).  Integrating over the
2200 \AA\ filter bandpass, this indicates a typical equivalent width
in the lobes of $\sim$4400$\pm$400 \AA, which is much higher than in
the central star because of the cooler and optically thinner dust
continuum (see below).

Most of the extended 11.9~$\micron$ emission -- and therefore most of
the dust mass -- appears to be concentrated in caps at the top and
bottom of the polar lobes, rather than in their side walls.  If the
dust mass traces the gas mass, then this is a crucial fact as it bears
on the latitudinal distribution of mass in the initial ejection.
There is good reason to think that the dust mass does generally trace
the total gas mass in the lobes, since the morphology of the 11.9
$\micron$ image is almost identical to the [Ne~{\sc ii}] images, as
well as narrowband emission-line images at shorter wavelengths (Smith
2003; Santander-Garcia et al.\ 2004; Guerrero et al.\ 2004).  On the
other hand, the regions of the polar lobes where the dust appears to
be deficient -- in the side walls at low latitudes and in the polar
``blisters'' beyond the lobes -- are precisely the locations of low
excitation where infrared [Fe~{\sc ii}] emission is mysteriously
enhanced (Smith 2003).  In other words, there is an anti-correlation
between dust and [Fe~{\sc ii}] emission, which might suggest that some
of the excess [Fe~{\sc ii}] emission has resulted as iron atoms were
liberated from grains into the gas phase by shocks.  We also point out
that the polar ``blisters'' are seen in the [Ne~{\sc ii}] image,
especially to the north.

Interestingly, the dust morphology resembles the high-excitation
emission tracers like He~{\sc i} and [O~{\sc iii}] somewhat better
than it does hydrogen lines.  For example, the missing side walls of
the polar lobes are a characteristic of both Figure 5$a$ and the
He~{\sc i} $\lambda$10830 emission (Smith 2003), while the
limb-brightened side walls are seen clearly in hydrogen and [Fe~{\sc
ii}] lines.  This is somewhat surprising, since dust and
high-excitation emission from ionzed gas are often segregated (the
nebula of RY Scuti is a clear example; Gehrz et al.\ 2001; Smith et
al.\ 2001).  The similarity of the warm dust emission and ionized gas
may have implications for the heating mechanism (i.e. either they are
heated radiatively by the same latitudinally-dependent UV field, or
perhaps the dust is heated by trapped Ly$\alpha$ photons in the
nebula, etc.).

Finally, several authors have noted excess extinction at low latitudes
(Smith 2003; Guerrero et al.\ 2004) or evidence for dust in the side
walls of the polar lobes from polarization data (Scarrott \& Scarrott
1995), while the majority of the dust in our images is located in the
polar caps as noted above.  This indicates that the extra dust that
may be causing this extinction and polarization in scattered light may
be cold, with insufficient heating to produce detectable mid-IR
emission.

\subsection{Dust Temperature and Optical Depth}

Figures 5$c$ and 5$d$ show the spatial distribution of dust color
temperature and emitting optical depth in the polar lobes of Mz~3.
The color temperature for each pixel in Figure 5$c$ is given by

\begin{equation}
T_c=\frac{14404 \, [ (1/\lambda_2) - (1/\lambda_1) ]}
{\ln\{[F_{\nu}(\lambda_1)/F_{\nu}(\lambda_2)]\,(\lambda_1/\lambda_2)^{\beta+3}\}}
\, {\rm K}
\end{equation}

\noindent where $F_{\nu}(\lambda_1)/F_{\nu}(\lambda_2)$ is the flux
ratio between the continuum at $\lambda_1$ (11.9~$\micron$) and
$\lambda_2$ (17~$\micron$), and $-\beta$ is the dust emissivity
exponent.  We have assumed that at these wavelengths the grains have
an emissivity proportional to $\lambda^{-1}$ (i.e.,
$\beta$=1).\footnote{Note that if this assumed emissivity is wrong,
then smaller values of $\beta$ would result in higher derived color
temperatures.}  To calculate this color temperature image, we clipped
the input 11.9 $\micron$ image at a level of 0.18 Jy arcsec$^{-2}$,
with lower values set to zero.  The resulting color temperature map
was then used to calculate the distribution of the emitting optical
depth $\tau$ of warm grains in the lobes of Mz~3 (note that $\tau$ is
not the {\it absorption} optical depth).  The emitting optical depth
at some wavelength is given by

\begin{equation}
\tau = - \ln \Big{[}1 - \frac{I_{\nu}}{B_{\nu}(T_c)} \Big{]}
\end{equation}

\noindent where $I_{\nu}$ is the specific intensity in an input image,
and $B_{\nu}(T_c)$ is the Planck function corresponding to a given
color temperature in each pixel of the $T_c$ image.

The apparent dust temperature distribution in Mz~3's polar lobes in
Figure 5$c$ is fairly uniform, dominated by temperatures of 100--130
K, while peak temperatures in the unresolved core region exceed 300 K.
Thus, the spatially-resolved temperature structure confirms our
interpretation of the SED in Figure 4$b$. Consequently, the ``cool''
mass estimates we list in Table 3 are reliable values for Mz~3's
nebula, without contamination by dust in the central disk.  The $\tau$
map for Mz~3 in Figure 5$d$ shows values that typically range from
0.01 to 0.05, so it is unlikely that a significant amount of mass is
hidden in optically thick clumps.  Only in the brightest part of the
polar lobes, roughly 8\arcsec\ northwest of the star, do we see
optical depths that exceed 0.1.

The observed grain color temperature and the apparent separation
from the central engine provide a consistency check for evaluating our
assumptions about distance, luminosity, and grain properties.  Most of
the dust in the polar lobes appears to reside at a separation of
$\sim$10\arcsec\ or $R$=10$^4$ AU if the distance to Mz~3 is roughly 1
kpc.  The most likely bolometric luminosity of the central engine is
of order 10$^4$ $L_{\sun}$ (Smith 2003).  Then, the grain temperature
should correspond to

\begin{equation}
T_c = 28 \, \Big{[}   
\frac{Q_{abs}}{Q_e}\,\frac{L}{10^4 L_{\sun}}\,
  \Big{(}\frac{R}{10^4 {\rm AU}}\Big{)}^{-2}
\Big{]}^{\frac{1}{4}} \, {\rm K}
\end{equation}

\noindent where $Q_{abs}/Q_e$ is the ratio of absorption to emission
efficiency for the grains.  For blackbodies (i.e., $Q_{abs}/Q_e$=1) we
would expect a dust temperature around 30 K, much lower than observed.
The higher observed temperature would require $Q_{abs}/Q_e\simeq$200,
which would be consistent with small graphite grains with radii of
$a\simeq$0.1~$\micron$ (Gilman 1974).  This is consistent with the
roughly $\lambda^{-1}$ wavelength-dependence that we assumed initially
to calculate the dust temperature.  Alternatively, if the grains are
larger than 0.1~$\micron$ (i.e. a somewhat smaller value of
$Q_{abs}/Q_e$), then the bolometric luminosity could be somewhat
larger or the distance smaller than we assumed.

\section{CONCLUSIONS}

We presented thermal-IR images of the bipolar planetary nebulae M~2-9,
Mz~3, and He~2-104, obtained at the IRTF and at ESO.  The distribution
of dust in these systems is of interest as each object is strongly
bipolar with a bright central star, and each is thought to be powered
by a symbiotic binary system.

Our mid-IR images allowed us to construct SEDs of the bright,
unresolved central engines separate from the surrounding nebulae.  In
each case, we find that a warm dust component of several hundred
degrees can account for the core emission (260, 320, and 320 K in
M~2-9, Mz~3, and He~2-104, respectively), while a cooler component
produces the more extended emission from the bipolar lobes (85, 110,
and 95 K in M~2-9, Mz~3, and He~2-104, respectively).  Hot dust at
temperatures up to $\sim$900 K is needed to fit the near-IR SEDs of
each source, but these additional components contribute negligible
mass.

In every case, the dust mass for the unresolved core region is orders
of magnitude less than the mass of the extended lobes, raising doubts
that the hypothetical disks in the core could have been responsible
for pinching the waists of the bipolar nebulae.  Assuming a gas:dust
mass ratio of 230 appropriate for graphite grains and solar
composition, we find masses of roughly 0.8 and 0.6 M$_{\sun}$ in the
nebulae of M~2-9 and Mz~3, respectively.  These relatively high masses
require that this material was donated by intermediate-mass progenitor
stars.  The mass of He~2-104's nebula is much lower, in accordance
with the fact that we detect no extended dust emission from its much
fainter nebula.

We detected extended dust emission around both M~2-9 and Mz~3, and in
both cases it resembled the distribution of ionized gas seen in
narrowband emission-line images.  Our images of Mz~3 have the highest
signal-to-noise in the extended polar lobes, and we show that the
fairly uniform color temperature derived from these images can explain
the 110~K dust component that dominates the far-IR spectral energy
distribution.  Thus, the mass derived from the cool 110~K component
traces material in the polar lobes.  In the case of Mz~3, the dust is
concentrated at high latitudes.

We also note possible evidence for grain destruction in shocks,
indicated by an anticorrelation between dust seen in our images and
[Fe~{\sc ii}] emission in previously-published near-IR images.  While
dust appears to be absent or weak in these regions (i.e. the polar
``blisters'' of Mz~3), 12.8 $\micron$ [Ne~{\sc ii}] emission is
detected.  Except for these regions with enhanced [Fe~{\sc ii}]
emission, the thermal-IR continuum from dust closely resembles the
distribution of ionzed gas.

Given the dust temperatures of the as-yet unresolved central sources,
where dust presumably resides in a circumstellar disk, each of these
bright objects may prove to be worthwhile targets for mid-IR
interferometric observations to resolve the disks.

\acknowledgments \scriptsize

N.S.\ was supported by NASA through grant HF-01166.01A from the Space
Telescope Science Institute, which is operated by the Association of
Universities for Research in Astronomy, Inc., under NASA contract
NAS5-26555.  R.D.G.\ was supported by NASA, the NSF, and the United
States Air Force.


\begin{deluxetable}{lcccc}
\tabletypesize{\scriptsize}
\tablecaption{IRTF/MIRLIN Observations of M~2-9}
\tablewidth{0pt}
\tablehead{
\colhead{Filter} &\colhead{$\lambda$} &\colhead{$\Delta\lambda$}
            &\colhead{Exp.\ Time} &\colhead{F$_{\nu}$\tablenotemark{a}} \\ 
 \colhead{} &\colhead{($\micron$)} &\colhead{($\micron$)} 
            &\colhead{(sec.)} &\colhead{(Jy)}
}
\startdata
N1		&8.81	&0.87	&2700	&30.3	\\
$[$S~{\sc iv}]	&10.5	&2\%	&5300	&32.8	\\
$[$Ne~{\sc ii}]	&12.8	&2\%	&3900	&41.1	\\
Qs		&17.9	&2.0	&6200	&41.5	\\
Q5		&24.5	&0.76	&2700	&38.1	\\
\enddata
\tablenotetext{a}{Central star in a 4$\arcsec$ diameter aperture.}
\end{deluxetable}

\begin{deluxetable}{lcccccc}
\tabletypesize{\scriptsize}
\tablecaption{TIMMI2 Observations of Mz~3 and He~2-104}
\tablewidth{0pt}
\tablehead{
\colhead{Filter} &\colhead{$\lambda$} &\colhead{$\Delta\lambda$}
            &\colhead{Exp.\ Time (Mz3)} &\colhead{F$_{\nu}$(Mz3)}
            &\colhead{Exp.\ Time (He2)} &\colhead{F$_{\nu}$(He2)} \\ 
 \colhead{} &\colhead{($\micron$)} &\colhead{($\micron$)} 
            &\colhead{(sec.)} &\colhead{(Jy)}
            &\colhead{(sec.)} &\colhead{(Jy)}
}
\startdata
N8.9		&8.7	&0.78	&720	&34.3	&2880	&8.1	\\
N11.9		&11.6	&1.20	&1200	&36.9	&2880	&8.8	\\
$[$Ne~{\sc ii}]	&12.8	&0.22	&1080	&45.8	&4320	&8.7	\\
Q1		&17.0	&0.80	&1200	&45.7	&\nodata&\nodata\\
\enddata
\tablecomments{F$_{\nu}$ for each source corresponds to the bright central
star measured through a 4$\arcsec$ diameter circular aperture.}
\end{deluxetable}

\begin{deluxetable}{llccc}
\tabletypesize{\scriptsize}
\tablecaption{IR Luminosity and Dust Mass}
\tablewidth{0pt}
\tablehead{
 \colhead{Parameter} &\colhead{units}
                     &\colhead{M~2-9} &\colhead{Mz~3} &\colhead{He~2-104}
}
\startdata
T$_c$ warm dust &K                        &260 &320  &320 \\ 
L warm dust &L$_{\odot}$\tablenotemark{a} &470 &462  &106 \\ 
M warm dust &M$_{\odot}$\tablenotemark{a} 
            &4.0$\times$10$^{-6}$ &1.1$\times$10$^{-6}$ &2.6$\times$10$^{-7}$ \\ 
M warm gas  &M$_{\odot}$\tablenotemark{a} 
            &9.2$\times$10$^{-4}$ &2.5$\times$10$^{-4}$ &6.0$\times$10$^{-5}$ \\ 
T$_c$ cool dust &K                        &85  &110  &95  \\ 
L cool dust &L$_{\odot}$\tablenotemark{a} &483 &1770 &29 \\ 
M cool dust &M$_{\odot}$\tablenotemark{a} 
            &3.4$\times$10$^{-3}$ &2.6$\times$10$^{-3}$ &1.0$\times$10$^{-4}$ \\ 
M cool gas  &M$_{\odot}$\tablenotemark{a} &0.78 &0.60 &0.02 \\ 
\enddata 
\tablenotetext{a}{At a distance of 1 kpc.}
\end{deluxetable}

\begin{figure}
\epsscale{0.9}
\plotone{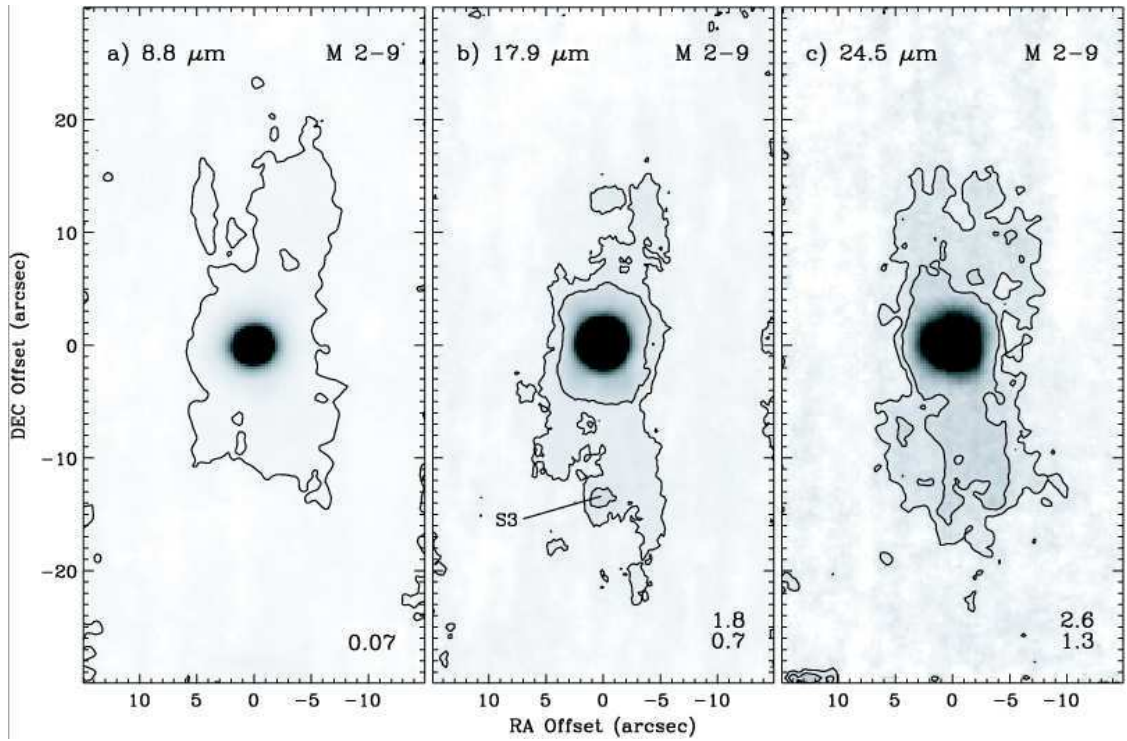}
\caption{IRTF/MIRLIN images of M~2-9 in continuum filters centered at
8.8 $\micron$ (a), 17.9 $\micron$ (b), and 24.5 $\micron$ (c).  Images
of the [S~{\sc iv}] and [Ne~{\sc ii}] emission lines obtained with the
CVF are not shown because no extended structure was detected.  The
lowest contour level is drawn at the 3$\sigma$ level above the
background, and contour levels in Jy arcsec$^{-2}$ are listed in the
lower right part of each frame.}
\end{figure}

\begin{figure}
\epsscale{0.9}
\plotone{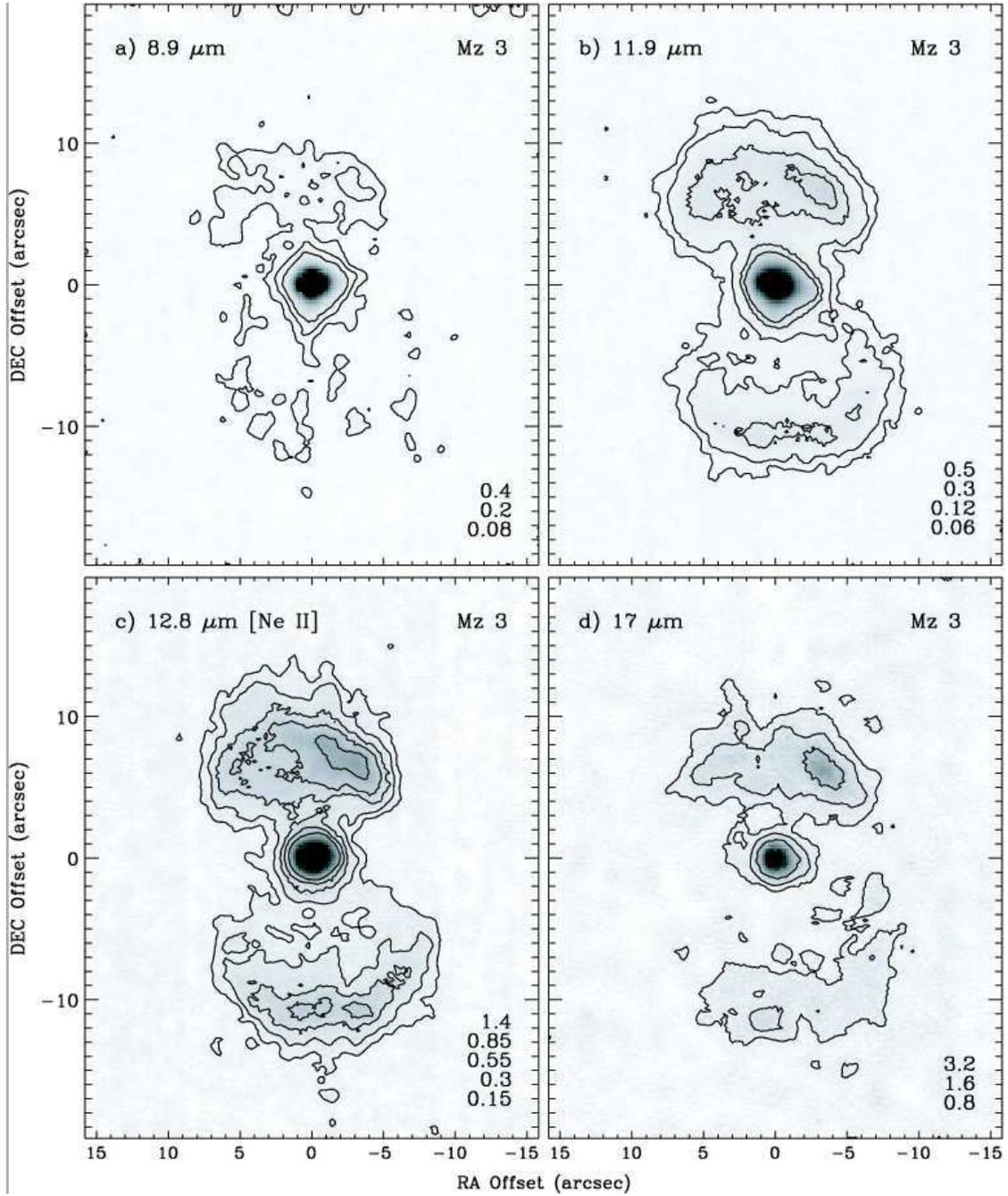}
\caption{TIMMI2 images of Mz~3 in the 8.9 $\micron$ continuum (a), in
the 11.9 $\micron$ continuum (b), in the 12.8 $\micron$ [Ne~{\sc ii}]
filter (c), and in the 17 $\micron$ continuum (d).  In each panel, the
lowest contour level is 3$\sigma$ above the background, and contour
levels in Jy arcsec$^{-2}$ are listed in the lower right.}
\end{figure}

\begin{figure}
\epsscale{0.99}
\plotone{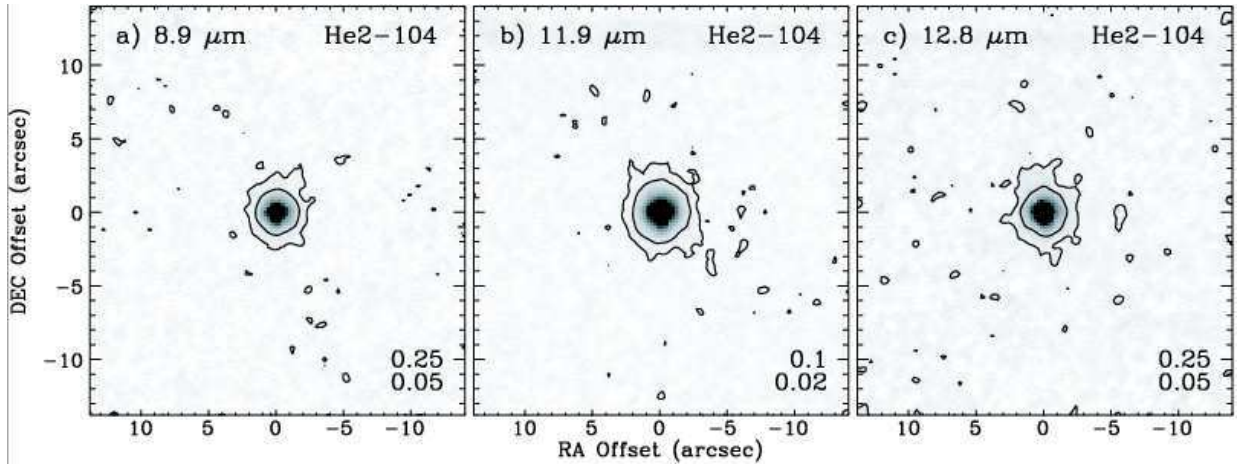}
\caption{TIMMI2 images of He~2-104 in continuum emission at 8.9
$\micron$ (a), 11.9 $\micron$ (b), and 17 $\micron$ (c).  In each
panel, the lowest contour level is 3$\sigma$ above the background, and
contour levels in Jy arcsec$^{-2}$ are listed in the lower right.}
\end{figure}

\begin{figure}
\epsscale{0.5}
\plotone{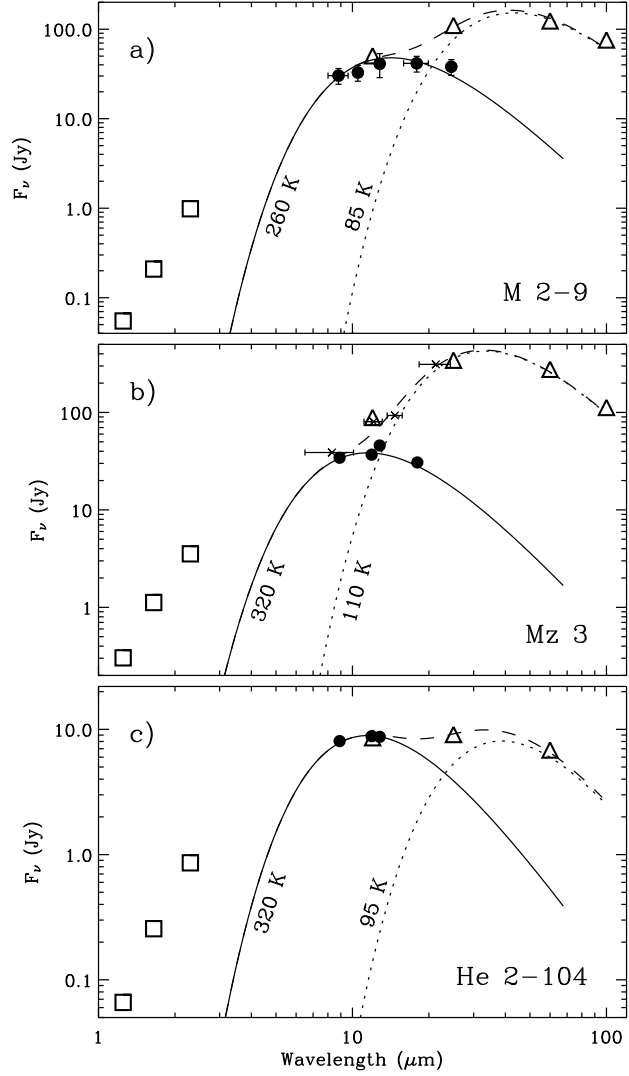}
\caption{Spectral energy distributions of M~2-9, Mz~3, and He~2-104 in
Panels a, b, and c, respectively.  Imaging photometry of the central
stars from Tables 1 and 2 is shown with solid circles, while 2MASS,
{\it MSX}, {\it IRAS} photometry are plotted with unfilled squares,
crosses, and unfilled triangles, respectively.  The solid line is an
approximate fit to the mid-IR photometry of the central star using
blackbodies with $\lambda^{-1}$ emissivity, the dotted line is a
similar fit to the integrated far-IR photometry, and the dashed line
shows the sum of both.  Warmer components are required to fit the
2MASS photometry, but these are not shown as they contribute
negligible mass.  The integrated fluxes including all extended
structure detected in our images of M~2-9 and Mz~3 are consistent with
the total {\it MSX} and {\it IRAS} fluxes at similar wavelengths, but
are not shown in this plot because of the large uncertainties.}
\end{figure}

\begin{figure}
\epsscale{0.9}
\plotone{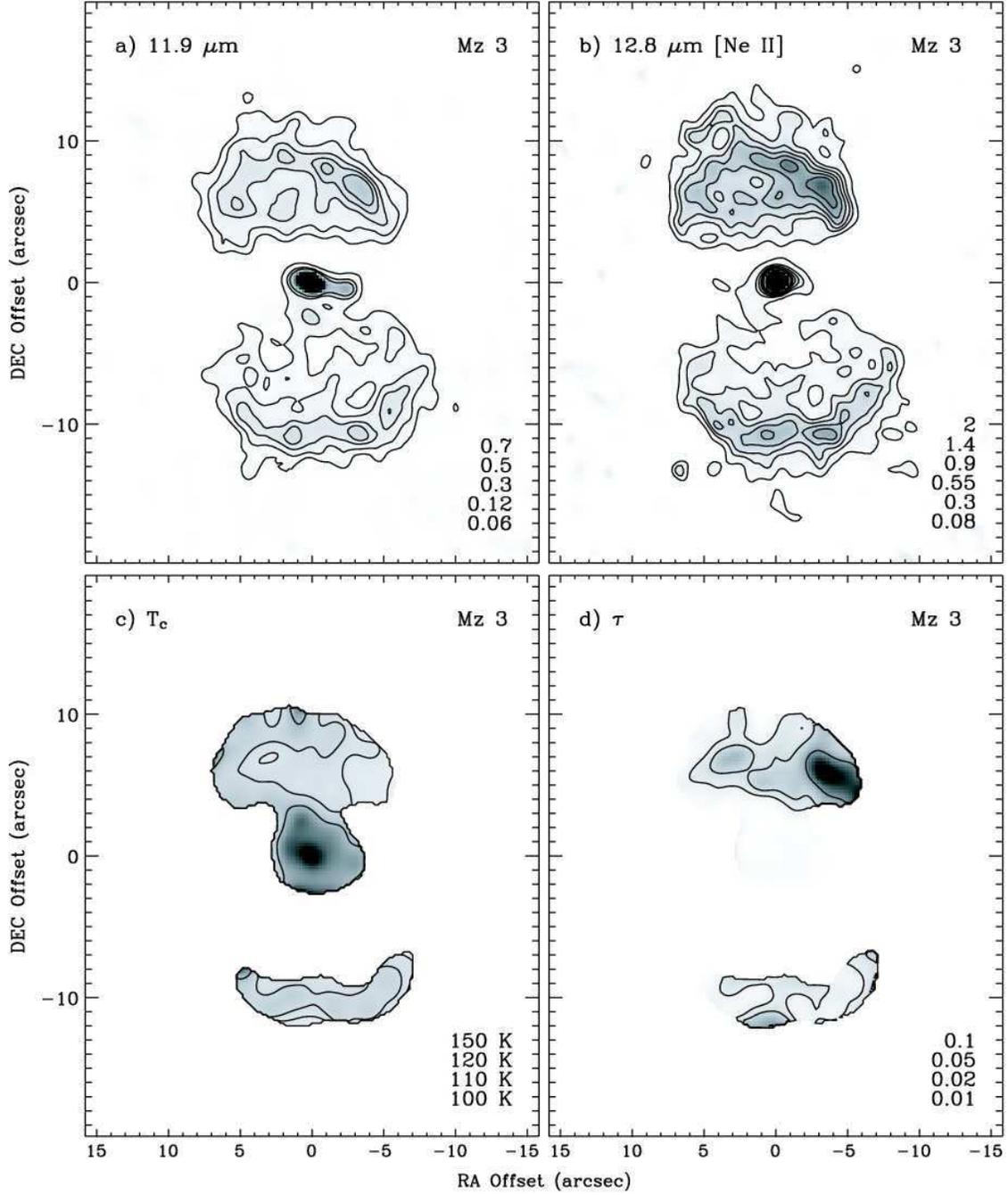}
\caption{Panels (a) and (b) show {\sc lucy}-deconvolved images of Mz~3
in the 11.9 $\micron$ continuum and in the 12.8 $\micron$ [Ne~{\sc
ii}] line.  Contour levels in Jy arcsec$^{-2}$ are listed in the lower
right corner of each panel.  (c) 11.9--17 $\micron$ color temperature
in the lobes of Mz~3.  The input 11.9 $\micron$ image was clipped at
0.18 Jy arcsec$^{-2}$ and all values below that set to zero, and the
11.9 $\micron$ image was smoothed to match the spatial resolution of
the 17 $\micron$ image.  Contours in Kelvin are listed in the lower
right.  (d) Optical depth image at 11.9 $\micron$ using the
temperature image in Panel (c), with optical depth contours listed in
the lower right.}
\end{figure}

\end{document}